\documentclass[12pt]{iopart}
\usepackage{iopams}

\usepackage{graphicx}
\usepackage{color}
\usepackage{bm}
\usepackage{epstopdf}
\usepackage{times}

\begin{document}

\title[Vibron-vibron coupling from AIMD]{Vibron-vibron coupling from \emph{ab initio} molecular dynamics simulations of a silicon cluster}

\author{Peng Han, Linas Vil\v{c}iauskas, Gabriel Bester}
\address{Max-Planck-Institut f\"ur Festk\"orperforschung, Heisenbergstra\ss e. 1, D-70569 Stuttgart, Germany}
\ead{g.bester@fkf.mpg.de}

\begin{abstract}
We study the temperature dependent dynamical processes of a Si$_{10}$H$_{16}$ cluster and obtain a blue shift of the Si-Si vibrational modes with transverse acoustic character and a red shift of the other vibrational modes with increasing temperature. We link this behavior to the bond length expansion and the varying sign of the Gr\"{u}neisen parameter. 
We further present a computational approach able to extract the vibron-vibron coupling strength in clusters or molecules. Our approach
is based on \emph{ab initio} Born-Oppenheimer molecular dynamics and a projection formalism able to deliver the individual vibron occupation numbers. From the Fourier transform of the vibron energy autocorrelation function we obtain the coupling strength of each vibron to the most strongly coupled vibronic states. We find vibron-vibron coupling strength up to 2.5 THz with a moderate increase of about 5\% when increasing the temperature from 50 to 150 K. 
\end{abstract}


\maketitle
\section{Introduction}
Colloidal semiconductor nanoclusters (NCs) have undergone an enormous development in the fields of optoelectronics, spintronics, and bio-labeling
over the past two decades. The good control of the NC size makes it possible to tailor their electronic and optical properties. Many successful applications in these fields were already reported~\cite{talapin10,pandey10,gerbasi10,baker10,mocatta11,viswanatha11,hetsch11}. The experiments are usually performed at room temperature, making the treatment of vibrational properties, especial the anharmonic effects, crucial~\cite{kraus07,pandey08,huxter10}.
A solid understanding of the anharmonic effects of vibrations, is therefore a decisive step for the real world application of nanostructures, where the physical properties such as thermal conductivity in nanowires~\cite{garg11}, non-radiative transition via phonon in NCs~\cite{han12c}, and Raman spectra broadening in nanostructures~\cite{doerk09} are dominated by the phonon lifetime.
There are basically two approaches to address this problem theoretically. 

The first one is to calculate the electron-phonon coupling~\cite{delerue01,vasilevskiy04,kelley11,zhang12}
and phonon-phonon coupling~\cite{shukla03,zibik04,turney09,lazzeri03,bonini07} terms via perturbative approaches.
The temperature effects are subsequently included using the Bose-Einstein distribution of the lattice vibrations (phonons).
The physical properties, such as spectral broadening, spin-flip, loss of quantum
coherence, relaxation of charge carriers, and thermal conductivity have been studied theoretically~\cite{zibik04,turney09,kelley11,zhang12,wei12},
mostly considering harmonic or quasi-harmonic (harmonic approximation performed at different volumes) approximations. 
Some have considered the third-order and fourth-order derivatives of the total energy~\cite{lazzeri03,bonini07}, 
but the large computational demand for these higher-order derivatives limit the study to very small nanostructures~\cite{martino09} or bulk materials~\cite{lazzeri03,bonini07}.
The $2n+1$ theorem, on the other hand limits the studies based on density functional perturbation theory (DFPT) to third order process~\cite{baroni01}.

The second one is to use molecular dynamics simulations~\cite{draeger03,gavartin06,franceschetti07,madrid09}, where the temperature effects including
all anharmonic effects are intrinsically contained in the atomic trajectories of the simulations.
The temperature-dependent vibrational density of states (DOS) and the thermal conductivity of nanostructures have already been studied using classical molecular dynamics simulations~\cite{henry08}. The required empirical force field potentials are limited by the lack of transferability to different systems
and by the inability to correctly predict chemical bonding processes. These difficulties are overcome in the case of \emph{ab initio} molecular dynamics (AIMD) 
simulations~\cite{gavartin06,franceschetti07,madrid09,kohli06,west07,turney09,hellman11}, which does not build upon the harmonic approximation
but allows for a self-consistent rearrangement of the changes, including all anharmonic effects.
Based on the accurately calculated forces from the electronic structure calculations, AIMD enables us to monitor the changes
in the electronic and vibrational properties on-the-fly, with thermal effects included directly.
Although AIMD simulations have been successfully applied to study a variety of problems~\cite{Marx09}, and the phonon lifetime in bulk has been studied using classical molecular dynamics simulations~\cite{ladd86,mcgaughey04,henry08}, a vibron-vibron interaction extracted from AIMD simulation is still lacking.

In this work, we perform AIMD simulations within the Born-Oppenheimer (BO) approximation of a Si$_{10}$H$_{16}$ cluster to study the geometry and the vibrational spectra 
from Fourier transformed velocity auto-correlation functions. The converged vibrational modes are obtained from a trajectory of about 46~ps (corresponding to 96,000 steps). We find a blue-shift of the Si-Si vibrational modes
with transverse acoustic (TA) characters and a red-shift of the other vibrational modes with increasing temperature. We also see a broadening of all the vibrational modes with increasing temperature. The former can be linked to the negative (blue-shift) and positive (red-shift) Gr\"{u}neisen parameters along with the extended bond lengths. The latter is attributed to the low symmetry (proximity to the surface) 
enhancing anharmonic effects at high temperatures. We further present a computational approach that enables the extraction of inter-vibron coupling strength, taking all the anharmonic effects into account. We find, for our cluster, couplings in the range of 0.15 to 2.5 THz which correspond to ``Rabi" oscillations of the vibron occupation number in the range of 0.4 to 6 ps. 

\section{Theoretical method}
\subsection{AIMD simulations}
\begin{figure}
\centerline{\includegraphics[width=4.0cm]{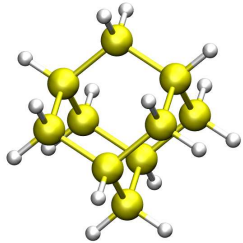}}
\caption{ 
Si$_{10}$H$_{16}$ cluster (sila-adamantane). Silicon atoms are drawn as large yellow spheres and hydrogen atoms as small white
spheres.
}\label{fig:Si10H16_geo}
\end{figure}

We construct a sila-adamantane (Si$_{10}$H$_{16}$) cluster that has the $T_d$ point group symmetry~\cite{lehtonen06,gaal08}, as shown in Figure~\ref{fig:Si10H16_geo}.
In this cage-shaped cluster, four silicon atoms are bonded to
three other silicon atoms and terminated by one hydrogen atom, while the remaining six silicon atoms are bonded to two silicon atoms and saturated with two hydrogen atoms.
The simulations are performed using density functional theory (DFT) within the local density approximation (LDA)
and Trouiller-Martin norm-conserving
pseudopotentials with an energy cutoff of 20~Ry~\cite{CPMD}.
The supercell is simple cubic with an extent of 15~\AA~ in each direction. The initial
geometry of the cluster is optimized until the minimum forces acting on the Si and H atoms are less than 3$\times$10$^{-6}$~a.u. under constrained symmetry. 

The AIMD simulations are initially performed at a constant temperature (NVT-ensemble) using the Nos\'{e}-Hoover chain thermostat~\cite{martyna92} 
with time step of 20~a.u. (about 0.48~fs) in order to equilibrating the system.
Following the 2~ps equilibration in the NVT-ensemble, we start a constant energy (NVE-ensemble) simulation and the trajectories are recorded at each time step.
All the AIMD simulations are performed with the CPMD code~\cite{CPMD}.
In order to improve the statistics, we chop the NVE-ensemble simulation into several slices. Each slice starts from a different NVT-ensemble equilibration time and
ends with the same number of NVE-ensemble simulation time steps.

\subsection{Vibrational DOS}

\begin{figure}
\centerline{\includegraphics[width=8.8cm]{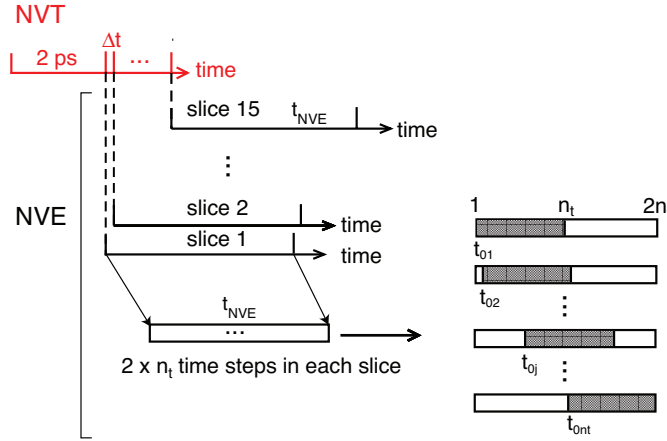}}\caption{ 
Schematic diagram of the AIMD simulation process. We start by an NVT simulation of 5.75 ps. From this simulation, we extract 15 initial configurations for the NVE runs. These are taken after an initial 2 ps of simulation time, with a time interval $\Delta t$ of 0.25 ps. We perform 15 NVE simulations (``slices'') lasting 3 ps (12 ps) for the calculation of vibrational DOS (vibrational cooling). For each slice we perform a simple moving average, as sketched on the lower right.
}
\label{fig:md_time}
\end{figure}

In order to obtain the temperature-dependent vibrational DOS from the AIMD simulation, we calculate
the velocity auto-correlation function $C(t)$~\cite{rahman64},
\begin{equation}
\label{eq:autocorrelation}
C(t) = \frac{\langle {\bm v}(t_0) \cdot {\bm v}(t_0+t) \rangle}{\langle {\bm v}^{2}(t_0) \rangle}
= \frac{\sum\limits_{k=1}^{n_s}\sum\limits_{j=1}^{n_t}\sum\limits_{i=1}^{N}{\bm v}_{i}(t_{0j}^k)\cdot {\bm v}_{i}(t_{0j}^k+t)}
{\sum\limits_{k=1}^{n_s}\sum\limits_{j=1}^{n_t}\sum\limits_{i=1}^{N} {\bm v}_{i}^{2}(t_{0j}^k)},
\end{equation}
where, $\langle ~~\rangle$ denotes the time averaged value calculated along the entire trajectory,  ${\bm v}_{i}(t_{0j}^{k})$ is the velocity vector of atom $i$ in slice $k$ at time origin point $j$. The number of atoms in the cluster, the number of time origin points, and the number of slices are labeled as $N$, $n_t$, and $n_s$, respectively. The time origin points labeled as $t_{01}$, $t_{02}$, $t_{0j}$ and $t_{0n_t}$ are given in Figure~\ref{fig:md_time}. In the present work, the number of slices $n_s$ is taken to be 15 with 6400 time steps and 3200 time origin points in each slice. The NVE-ensemble simulations are performed for a total of 96000 time steps corresponding to about 46~ps simulation time. One slice corresponds to a simulation time of approximately 3~ps. 

\begin{figure}
\centerline{\includegraphics[width=12.0cm]{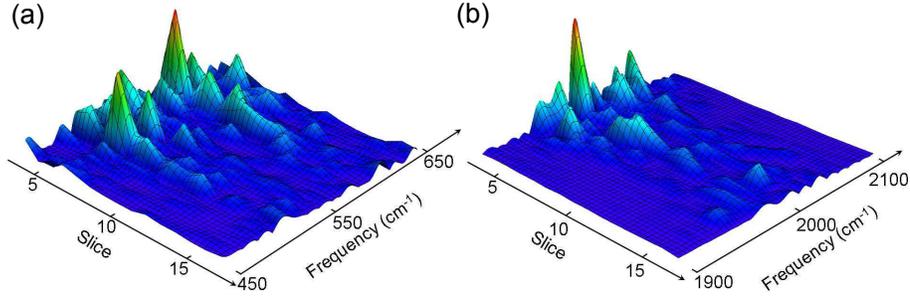}} \caption{
Convergence of the vibrational modes $\Delta VDOS_{k}(\omega)$ with frequency between (a) 450 and 650~cm$^{-1}$ (bending Si-H and rotation H-Si-H modes), and (b) 1900 and 2100~cm$^{-1}$
(stretching Si-H modes) at $T=$ 800~K as function of the simulation time given in unit of slices (3~ps). The vertical axis describes the deviation of the vibrational DOS between a simulation with $k$ and a simulation with $k-1$ time slices. The red color (maximum deviations) reflects a deviation of 30~\%.
}\label{fig:freq_conv}
\end{figure}

The vibrational DOS are calculated using the Fourier transform~\cite{rahman64}
\begin{equation}
\label{eq:ft}
VDOS(\omega) = \frac{1}{\sqrt{2\pi}}\int_{-\infty}^{+\infty} C(t) e^{-i\omega t} dt.
\end{equation}
and for demonstration of the convergence with the number of slices we use the change of the vibrational DOS 
\begin{equation}
\label{eq:slice}
\Delta VDOS_{k}(\omega) = |VDOS_{k}(\omega) - VDOS_{k-1}(\omega)| \quad ,
\end{equation}
showing the deviation of the vibrational DOS between a simulation with $k$ and a simulation with $k-1$ time slices.
Figure~\ref{fig:freq_conv} (a) and (b) shows $\Delta VDOS_{k}(\omega)$ for the low (a) and high (b) frequency range at a temperature of 800 K. In this figure, the maximum deviation is 30\%. We observe that the vibrational modes start to converge after 35 ps (12 slices).

Since the limited statistic we obtained from AIMD at low temperature even for a small cluster, we use the DFPT results based on the harmonic approximation of lattice dynamics as a low temperature limit. 
In this case, the DFPT represents a very good classical (neglecting zero-point motion) approximation.
The vibrational frequencies $\omega$ and the vibrational eigenvectors ${\bm U_{i}}$ are obtained from the eigenvalue equation:~\cite{Yu10,Madelung96}
\begin{equation}
\label{eq:eigen}
\sum_{j=1}^{N}\frac{1}{\sqrt{M_{i}M_{j}}}\frac{\partial^{2}V}{\partial{\bm R_{i}}\partial{\bm R_{j}}}
{\bm U_{j}}=\omega^{2}{\bm U_{i}},
\end{equation}
where, $M_{i (j)}$ is the mass of atom $i$ ($j$), $V$ is the potential, $\bm R_{i (j)}$ denotes the atomic positions.
The dynamical matrix elements $\frac{1}{\sqrt{M_{i}M_{j}}}\frac{\partial^{2}V}{\partial{\bm R_{i}}\partial{\bm R_{j}}}$ are calculated using linear response as implemented in the CPMD code~\cite{CPMD}. In this approach, the second derivative of the potential is computed from the linear response of the system to an infinitesimal displacement based on perturbation theory~\cite{baroni01}.

\subsection{Energy relaxation}

The extraction of  the potential energy of a certain vibrational mode, $E_p^{\nu}(t)$, was first proposed by Ladd \emph{et al.}~\cite{ladd86} and subsequently modified by McGaughey \emph{et al.}~\cite{mcgaughey04} to include the total energy of each mode instead of the potential energy only~\cite{mcgaughey04,henry09}. In the quasi-harmonic approximation, where the changes in bond length due to thermal expansion are included but further anharmonic effects are excluded~\cite{turney09}, the total energy of each vibrational mode $E^{\nu}(t)$ ---proportional to the occupation number of the vibrational mode--- is expressed in terms of the time-dependent normal coordinates $Q_{\nu}(t)$
\begin{equation}
\label{eq:Etotal}
E^{\nu}(t)=\frac{1}{2}[\dot{Q^{*}_{\nu}}(t)\dot{Q_{\nu}}(t)+\omega_{\nu}^{2}Q^{*}_{\nu}(t)Q_{\nu}(t)],
\end{equation}
where
\begin{equation}
\label{eq:Qmode}
Q_{\nu}(t)=\sum_{i}^{N}\sqrt{\frac{M_i}{N}}{\bm U^{\nu}_{i}}\cdot[{\bm R_{i}}(t)-{\bm R_{i}^{0}}].
\end{equation}
The three-component vectors ${\bm U^{\nu}_{i}}$ in Eq.~(\ref{eq:Qmode}) represent the three components belonging to atom $i$ of the vibrational eigenvectors obtained from Eq.~(\ref{eq:eigen}).
${\bm R_{i}^{0}}$ is the equilibrium position of atom $i$ obtained from the trajectory of the AIMD simulation as,
\begin{equation}
\label{eq:Atomposi}
\bm R_{i}^{0} = \frac{1}{n_tn_s}\sum_{k=1}^{n_s}\sum_{j=1}^{n_t} \bm R_{i} (t^{k}_{j}).
\end{equation}
Based on the quasi-harmonic approximation, 
the vibrational vectors ${\bm U^{\nu}_{i}}$ used in Eq.~(\ref{eq:Qmode}) are calculated using DFPT with the atomic positions obtained from Eq.~(\ref{eq:Atomposi}),
i.e. from an AIMD simulation at a certain temperature.

The attenuation of the vibrational amplitude reflects the mode relaxation processes and can be described by the auto-correlation function of the energy fluctuation, written as
\begin{equation}
\label{eq:E_relax}
C^\nu_E(t) = \frac{\langle \delta E^{\nu}(t_0) \delta E^{\nu}(t_0+t)\rangle}{\langle \delta E^{\nu}(t_0) \delta E^{\nu}(t_0)\rangle}
\end{equation}
where, $\langle ~~\rangle$ denotes the time averaged value calculated along the entire trajectory and $\delta E^{\nu}(t)=E^{\nu}(t)-\overline{E^{\nu}}$ is the deviation from
the average vibrational energy $\overline{E^{\nu}}$. 

\section{Temperature dependent vibrational properties}\label{sec:vdos}

\begin{figure}
\centerline{\includegraphics[width=7.0cm]{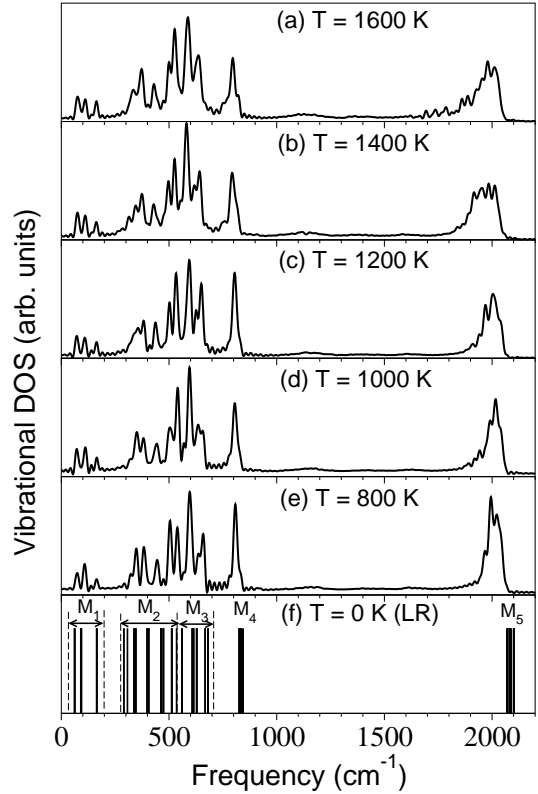}} \caption{
Vibrational DOS of a Si$_{10}$H$_{16}$ cluster obtained from AIMD simulations at the
temperatures of (a) 1600~K, (b) 1400~K, (c) 1200~K, (d) 1000~K, and (e) 800~K. The vibrational modes of a
Si$_{10}$H$_{16}$ cluster obtained from linear-response (LR) calculations at zero temperature is given
in (f) including the mode assignment. 
}\label{fig:si10H16_vdos}
\end{figure}

We now describe the temperature dependence of the  vibrational properties.
In Figure~\ref{fig:si10H16_vdos}, we plot the converged 
vibrational DOS calculated using Eq.~(\ref{eq:autocorrelation}) and (\ref{eq:ft}) at 
(a) 1600~K, (b) 1400~K, (c) 1200~K, (d) 1000~K, and (e) 800~K along with (f) the zero temperature linear-response results from Eq.~(\ref{eq:eigen})
(all ignore the zero point vibrations). 
In this work, we use the linear-response results as a low temperature limit because of the limited
statistic we obtained from AIMD at low temperature even for the small cluster.
The vibrational eigenmodes obtained from the linear-response calculations are analyzed in terms of their displacement pattern by a projection onto bulk
phonon modes (see Ref.~\cite{han12}) and are divided into acoustic-like Si-Si modes and optical-like Si-Si modes.
The vibrations between the silicon and hydrogen atoms are divided into 
bending Si-H and rotation H-Si-H modes, shear H-Si-H modes and stretching Si-H modes according to the displacement of the atoms. 
These vibrational modes are listed in Table~\ref{table:vmod} and labeled as M$_1$-M$_5$ in Figure~\ref{fig:si10H16_vdos} (f).
\begin{table}
\caption{Vibrational modes of Si$_{10}$H$_{16}$ cluster.}
\label{table:vmod}
\begin{tabular}{lcc}
\hline
\hline
                           & label  ~~~~~~~& frequency (cm$^{-1}$) \\
\hline
acoustic-like Si-Si modes   & $M_1$  &  $<$  200                \\
optical-like  Si-Si modes   & $M_2$  &  300 $\rightarrow$ 550           \\
bending Si-H and rotation H-Si-H modes  & $M_3$  & 550 $\rightarrow$ 700\\
shear H-Si-H modes          & $M_4$  &  800 $\rightarrow$ 900           \\
stretching Si-H modes       & $M_5$  &   $\approx$ 2100              \\ 
\hline
\hline
\end{tabular}
\end{table}
In the AIMD simulations, the character of the vibrational eigenmodes cannot be obtained directly~\cite{martinez06}.

From Figure~\ref{fig:si10H16_vdos}, we observe that the vibrational DOS of all the modes, especially for the Si-H modes, show a broadening with increasing temperature and that the vibrational density of the shear H-Si-H modes and the stretching Si-H modes decreases with increasing temperature. There is also a red shift with increasing temperature of all the vibrations except for acoustic-like Si-Si modes, which correspond to the TA phonon modes of bulk Si.

\begin{figure}
\centerline{\includegraphics[width=8.5cm]{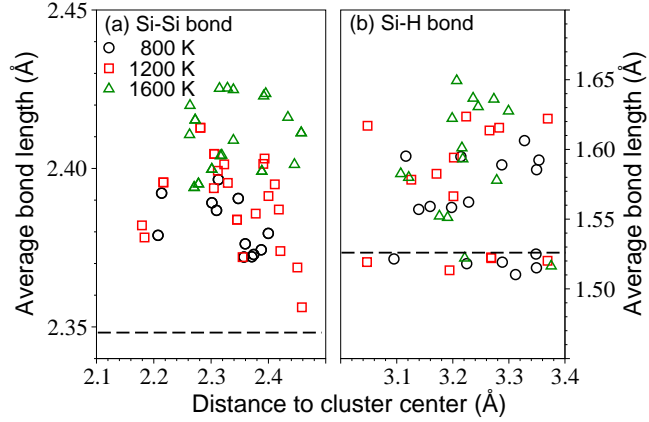}} \caption{
Thermally averaged bond length distribution of (a) Si-Si bond and (b) Si-H bond as a function of their distance (bond center) to
the dot center at $T=$ 800~K (black circles), 1200~K (red squares),
and 1600~K (green triangles). The optimized bond lengths of the Si$_{10}$H$_{16}$ cluster
at $T=$ 0~K are given as dashed lines.
}\label{fig:Si10H16_bondlength}
\end{figure}

To explain the reason for this temperature dependence, we plot in Figure~\ref{fig:Si10H16_bondlength} the average bond lengths of (a) Si-Si bonds
and (b) Si-H bonds as a function of the distance of the respective bond center to the cluster center at $T=$ 800, 1200, and 1600~K. In the cage-shaped Si$_{10}$H$_{16}$ cluster,
the bond lengths of each Si-Si bond (Si-H bond) are uniform after geometry optimization at zero temperature. We plot these optimized bond lengths as dashed lines
in Figure~\ref{fig:Si10H16_bondlength}. We see from Figure~\ref{fig:Si10H16_bondlength} (a) that the Si-Si bond lengths increase with increasing temperature. Moreover, the cluster shows a large bond length distribution at high temperatures. The increase of bond length and its large distribution along with the positive Gr\"{u}neisen parameters (describing the change in phonon frequencies with volume) for the optical branches and longitudinal acoustic (LA) branch result in the softening (red shifting) and broadening of the optical-like and the LA-like Si-Si modes with increasing temperature. In contrast to the relatively small extension of Si-Si bonds ($\approx$ 3$\%$), the Si-H bonds show a large extension and distribution with increasing temperature. A bond length extension of as much as 9$\%$ at $T=$ 1600~K is seen in Figure~\ref{fig:Si10H16_bondlength} (b). We attribute the large extension of the Si-H bonds to the geometry of the Si$_{10}$H$_{16}$ cluster. In contrast to the silicon atoms, which are localized at the center of the tetrahedral structure (see Figure~\ref{fig:Si10H16_geo}), the hydrogen atoms at the cluster surface are free to move outwards. This introduces a potential asymmetry towards the vacuum side and increases the anharmonicity.
A relatively large extension of Si-H bonds is therefore obtained at high temperature, which explains the red shift and the broadening of the  Si-H vibrational modes, considering the positive Gr\"{u}neisen parameters. Especially, the vibrations of the high frequency H-Si-H shear and Si-H stretching modes mainly consist of the displacements of hydrogen atoms. This leads to a significant reduction and broadening of the vibrational peaks corresponding to the H-Si-H shear modes and Si-H stretching modes in Figure~\ref{fig:si10H16_vdos} (a)-(e).
Finally, the blue shift of the TA-like Si-Si modes observed in Figure~\ref{fig:si10H16_vdos} is the result of the bond length extension along with the 
negative Gr\"{u}neisen parameters of the TA modes.

\section{Vibron-Vibron Coupling}

\begin{figure}
\centerline{\includegraphics[width=8.0cm]{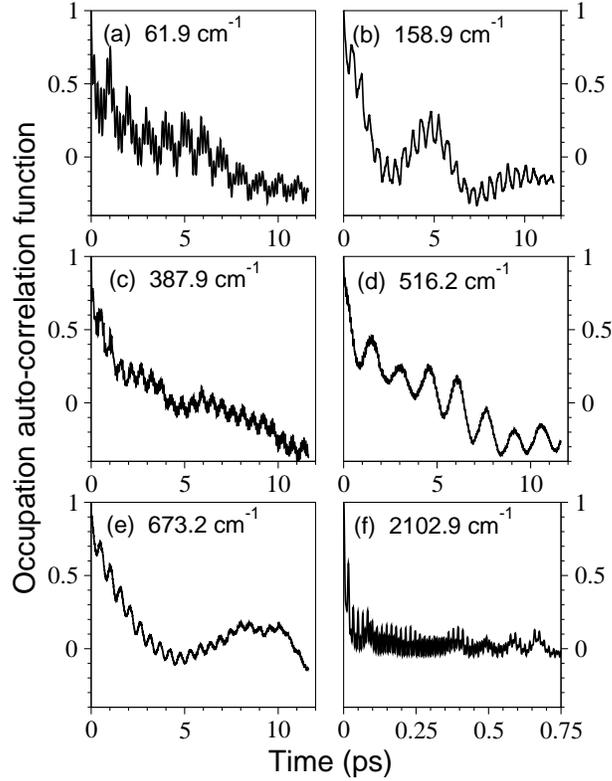}} \caption{
Occupation autocorrelation functions of vibrational modes with frequencies (a) 61.87~cm$^{-1}$, (b) 158.85~cm$^{-1}$, (c) 339.52~cm$^{-1}$, (d) 516.23~cm$^{-1}$, (e) 673.16~cm$^{-1}$, and (f) 2102.95~cm$^{-1}$ at $T=$ 100~K.
}\label{fig:Si10H16_moderelax100k}
\end{figure}

The vibrational energy of a certain mode is proportional to the mode occupation number $E^{\nu}(t) \propto n^{\nu}(t)$ (see Eq.~\ref{eq:Etotal}), and thus the time evolution of $E^{\nu}(t)$ carries the information about the coupling between different vibrational modes. In the following we will refer to $E^{\nu}(t)$  as to the ``occupation auto-correlation function'', to avoid confusions with the many variants of the energy auto-correlation function. The total vibrational enegy, $\sum_\nu E^\nu$, is conserved in our NVE simulation and energy is allowed to flow between vibrational modes. To extract a meaningful statistical average out of $E^{\nu}(t)$ we use the correlation function $C^\nu_E(t)$ (Eq.~\ref{eq:E_relax}), which decays to zero for large times $t$, when the occupation number is unrelated to the initial situation (at time $t_0$). We have calculated the occupation auto-correlation functions for all the vibrational modes of the Si$_{10}$H$_{16}$ cluster at $T=$ 50, 100, 125, and 150~K. In figure~\ref{fig:Si10H16_moderelax100k} (a)-(f) we plot the vibrational occupation autocorrelation function of six selected vibrational modes of the Si$_{10}$H$_{16}$ cluster at $T=$ 100~K, which will be discussed in more detail. 

It should be pointed out that the AIMD simulations in this work are performed only at relatively low temperatures in order not to violate the quasi-harmonic approximation which is used to estimate the vibrational eigenmode energy~\cite{ladd86,mcgaughey04}. Indeed, the vibrational vectors ${\bm U^{\nu}_{i}}$ are calculated using DFPT with the average finite-temperature atomic positions obtained from Eq.~(\ref{eq:Atomposi}). Imaginary vibrational frequencies are introduced when the atoms deviate far from their equilibrium positions at high temperatures.

First, we have to prominently assert that a lifetime, as can be extracted from a damped Rabi oscillation or from transitions treated at the level of Fermi's golden rule does not exist in our cluster. The NVE simulation does not allow for energy dissipation, that would possibly lead to exponential decay of high energy phonon modes, from which a lifetime could be extracted. On the other hand, the discrete and energetically sparse nature of the vibrons does not allow for a treatment following Fermi's golden rule, where transitions into a perfectly dissipative continuum are assumed. Our occupation numbers describe the time evolution of a many vibronic level system without dissipative term. The change of occupation of the individual levels reflects the interlevel coupling. The strongest coupling will dominate the short time evolution.

\begin{figure}
\centerline{\includegraphics[width=8.0cm]{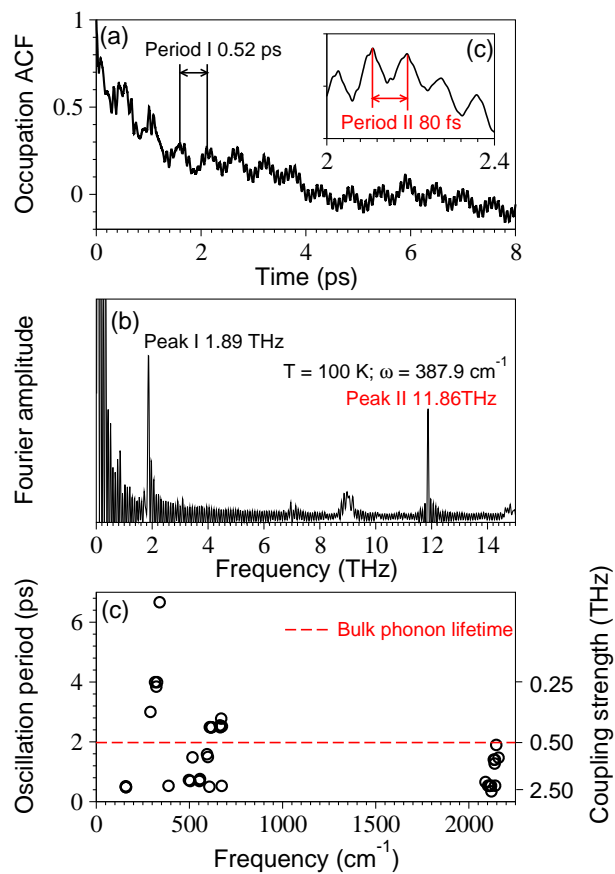}} \caption{
(a) Occupation autocorrelation function of one vibrational mode (see Figure~\ref{fig:Si10H16_moderelax100k} (c)) with labeled vibron-vibron oscillation period (Period I) and vibrational oscillation period (Period II). (b) Fourier transform of the occupation autocorrelation function of (a) vibrational mode with labeled Peak I corresponding to the vibron-vibron coupling strength and peak II coresponding to the vibrational frequency. (c) The vibron-vibron oscillation period for all the vibrational modes (circles) at 100~K along with the bulk phonon lifetime (dashed line) from experiment.
}\label{fig:ft_analyze}
\end{figure}

In figure~\ref{fig:Si10H16_moderelax100k} (a)-(f) we observe a rather different behavior of the different vibrational modes. In all cases, however, there is an ultrafast oscillation superimposed on a slower variation. To analyze  these results in a quantitative way, we replot in figure~\ref{fig:ft_analyze} (a) the occupation autocorrelation function for the vibrational mode with frequency 387.9 cm$^{-1}$ (Fig.~\ref{fig:Si10H16_moderelax100k} (c)) along with its Fourier transform (panel b). The Fourier analysis shows that the high frequency oscillation (Period II in \ref{fig:ft_analyze} (a)), in this case at 11.86 THz, corresponds to the vibron frequency. The origin of this oscillation resides in errors in the projection (Eq.~(\ref{eq:Qmode})). Indeed, we project our AIMD anharmonic results onto the harmonic vibrations of the NC. This represents an approximation with an error that is largest when the atomic displacements have the largest amplitude. Accordingly, the error increases and decreases periodically with a frequency equal to the vibrational mode's frequency.  The slower variation (Period I in \ref{fig:ft_analyze} (a)) is very clear from Fig.~\ref{fig:ft_analyze} (b) and has a frequency of 1.89 THz. We identify this feature  as the frequency with which the vibron state undergoes periodic oscillations with strongly coupled vibronic states. It describes how the specific vibron mode can transfer energy to other modes, i.e., decay into lower energy vibrations or how two, or more, low energy vibrations can create a high energy vibron. Both processes are present in our  AIMD calculations. The fact that we are dealing with a many-level (we have 78 vibrational eigenmodes) system explain that we do not see clear Rabi oscillations between two distinct levels but a fast oscillation (1.89 THz) modulated in time by the interaction with more weakly coupled levels. We cannot extract the weak coupling components from our results (that are all hidden in the low frequency Fourier transform Figure~\ref{fig:ft_analyze} (b), and would require much longer simulation times to be properly captured), but can clearly obtain the strongest coupling. 

In Figure~\ref{fig:ft_analyze} (c) we summarize our Fourier analysis of all the vibron modes and plot the oscillation period as a function of the vibron frequency. We find strongly coupled (short oscillation periods) modes at all available frequencies, starting from the bulk acoustic-like modes to the passivant hydrogen vibrations. The coupling strength varies between 0.15 and 2.5 THz. 

\begin{figure}
\centerline{\includegraphics[width=8.0cm]{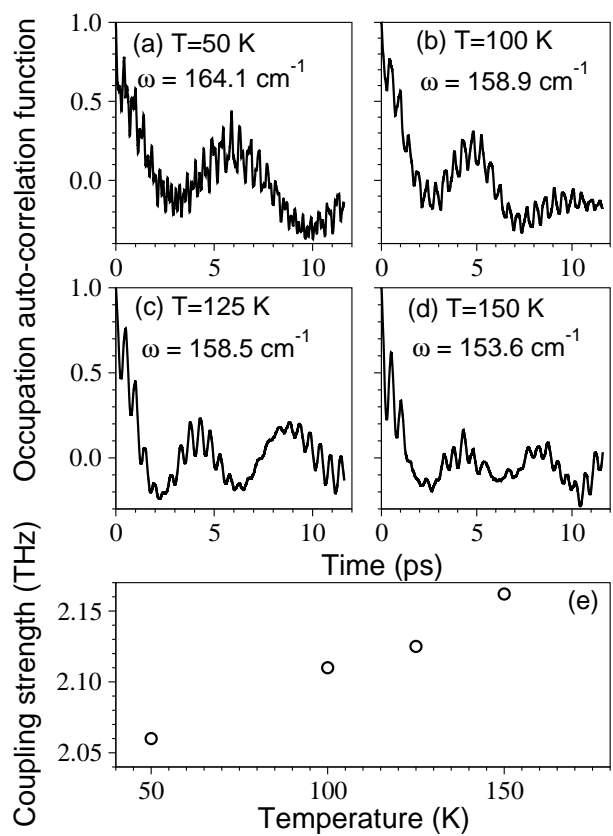}} \caption{
Occupation autocorrelation functions of a certain vibrational mode (see Figure.~\ref{fig:Si10H16_moderelax100k} (b)) at temperature (a) 50~K, (b) 100~K, (c) 125~K, and (d) 150K. The vibron-vibron coupling strength obtained from the occupation autocorrelation functions are given in (e).
}\label{fig:Si10H16_moderelaxtemperture}
\end{figure}

We now study the temperature effects on the coupling strength of a certain vibrational mode.
In Figure~\ref{fig:Si10H16_moderelaxtemperture} (a)-(d), we plot the occupation autocorrelation functions of a certain acoustic-like mode with varying temperatures and in Fig.~\ref{fig:Si10H16_moderelaxtemperture}(e) the extracted coupling strengths. We find that the vibrational frequencies show a red shift with increasing temperature which can be related to the thermal expansion and behaves as expected, and that the coupling strength increases slightly, from 2.06 to 2.16 THz. 
This increase can be attributed to higher-order anharmonic effects~\cite{srivastava80,henry08}. With the increase in temperature, 
the effects of higher order anharmonicity, which includes four-vibron, five-vibron, and even higher order interactions, become stronger~\cite{west07}. 

\section{Summary}

We performed first-principles AIMD simulations, within the BO approximation, to study the temperature dependent vibrational properties of a Si$_{10}$H$_{16}$ cluster. We first calculate the vibrational DOS using the Fourier transformed velocity autocorrelation functions and compare the results to DFPT, obtaining good agreement. We quantify the softening and broadening, with increasing temperature, of the Si-Si LA-like modes, the Si-Si optical modes and the Si-H modes (especially for the H-Si-H shear modes and Si-H stretching modes) and analyze the results in terms of bond length variations. Subsequently, we suggest a method to calculate the vibron-vibron coupling strength, including all the anharmonic effects, using a Fourier transformation of the occupation autocorrelation function. For our Si cluster, we find a coupling strength between 0.15 and 2.5 THz, which corresponds to oscillations in the occupation of vibron states (akin Rabi oscillations) in the range of 0.4 to 6 ps. The obtained coupling parameters could be used in further study of the dynamical processes in nanosystem based on rate equations and potentially including dissipation through the coupling to an external phonon bath \cite{romano07,galperin07}. We conclude that, although our approach enables the extraction of anharmonic vibron coupling parameters otherwise not available, it still requires trajectories of around 24 ps, which presently limits it's applicability to rather small structures. Further challenges involve the non-adiabatic coupling to electronic states and the coupling to a dissipative environment, which we have ignored.

\section*{Acknowledgments} 
We would like to acknowledge financial support by the BMBF (QuaHL-Rep, Contract
No. 01BQ1034). Most of the simulations were preformed on the national supercomputer NEC Nehalem Cluster
at the High Performance Computing Center Stuttgart (HLRS).

\end{document}